\documentclass{article}

\PassOptionsToPackage{numbers}{natbib}

\usepackage[preprint]{neurips_2025}

\usepackage[utf8]{inputenc} 
\usepackage[T1]{fontenc}    
\usepackage{hyperref}       
\usepackage{url}            
\usepackage{booktabs}       
\usepackage{amsfonts}       
\usepackage{nicefrac}       
\usepackage{microtype}      
\usepackage{xcolor}         
\usepackage{amsmath}
\usepackage{amssymb}
\usepackage[capitalise, noabbrev]{cleveref}
\usepackage{bm}
\usepackage{enumitem}
\usepackage{xspace}
\usepackage{multirow}
\usepackage{graphicx}
\usepackage{float} 
\usepackage{subfigure}
\usepackage[ruled,linesnumbered]{algorithm2e}
\usepackage{algpseudocode}
\usepackage{makecell}
\usepackage{hyperref}
\usepackage{url}
\usepackage{dsfont}
\usepackage{colortbl}
\usepackage{tikz}
\usepackage{setspace}
\usepackage{wrapfig}
\usepackage{tcolorbox}

\newtheorem{defn}{\bf Definition}
\newtheorem{prbl}{\bf Problem}
\newtheorem{prop}{\bf Proposition}

\newcommand{\ie}{\textit{i.e.,}\xspace}

\newcommand{\eg}{\textit{e.g.,}\xspace}
\newcommand{\cmt}[1]{\textcolor{purple}{[#1]}}
\newcommand{\rev}[1]{\textcolor{black}{#1}}

\newcommand{\ir}[1]{\textcolor{black}{#1}}
\newcommand{\model}{LLMFOSA\xspace}
\newcommand{\ml}{ML-1M\xspace}
\newcommand{\lastfm}{LFM-360K\xspace}
\newcommand{\rep}[1]{\bm{#1}}

\title{Improving Recommendation Fairness without Sensitive Attributes Using Multi-Persona LLMs}

\author{
  Haoran Xin$^1$ \quad Ying Sun$^1$ \quad Chao Wang$^2$ \quad Yanke Yu$^1$ \quad Weijia Zhang$^1$ \quad Hui Xiong$^1$ \\
  $^1$ Thrust of Artificial Intelligence, The Hong Kong University of Science and\\
  Technology (Guangzhou), $^2$ School of Artificial Intelligence and Data Science, \\
  University of Science and
  Technology of China \\
  \texttt{\{hxin883, wzhang411\}@connect.hkust-gz.edu.cn, yings@hkust-gz.edu.cn} \\
  \texttt{wangchaoai@ustc.edu.cn,}
  \texttt{rankyurky@gmail.com, xionghui@ust.hk}
}

\begin{document}

\maketitle

\begin{abstract}
Despite the success of recommender systems in alleviating information overload, fairness issues have raised concerns in recent years,
potentially leading to unequal treatment for certain user groups.
While efforts have been made to improve recommendation fairness, they often assume that users' sensitive attributes are available during model training.
However, collecting sensitive information can be difficult, especially on platforms that involve no personal information disclosure.
Therefore, we aim to improve recommendation fairness without any access to sensitive attributes.
This is a non-trivial task because uncovering latent sensitive patterns from complicated user behaviors without explicit sensitive attributes can be difficult.
Consequently, suboptimal estimates of sensitive distributions can hinder the fairness training process.
To address these challenges, leveraging the remarkable reasoning abilities of Large Language Models~(LLMs), we propose a novel \underline{LLM}-enhanced framework for \underline{F}air recommendation with\underline{O}ut \underline{S}ensitive \underline{A}ttributes (\model).
A Multi-Persona Sensitive Information Inference module employs
LLMs with distinct personas that mimic diverse human perceptions to infer and distill sensitive information.
Furthermore, a Confusion-Aware Sensitive Representation Learning module incorporates inference results and rationales to develop robust sensitive representations, considering the mislabeling confusion and collective consensus among agents.
The model is then optimized by a formulated mutual information objective.
Extensive experiments on two public datasets validate the effectiveness of \model in improving fairness.

\end{abstract}

\section{Introduction}
\vspace{-3mm}
In the era of information explosion, recommender systems (RSs) play a crucial role in people's daily lives, helping users discover content they might enjoy from vast amounts of information.
Despite their significant success, concerns about fairness in RSs have grown in recent years~\cite{wang2023survey,li2021towards,zhao2023fair,wu2021learning}.
\rev{The unfairness primarily stems from training on data that implicitly contains biased behaviors correlated with sensitive information~(\eg gender, race),
even if such information is not explicitly available.%
}
Unfairness can lead to unequal treatment for certain user groups, potentially harming user experience and the platform's overall interests.
For instance, on job recruitment platforms, female users might receive fewer recommendations for high-paying jobs compared to their male counterparts~\cite{lambrecht2019algorithmic}.

Extensive research efforts have been made to improve fairness in RSs.
Some works \cite{yao2017beyond,aghaei2019learning} directly employed fairness metrics as regularization terms to balance recommendations among different user groups.
Another line of research \cite{bose2019compositional,zhang2023fairlisa,wu2021learning,zhang2023fairlisa,dai2021say} applied adversarial learning to train filters to remove hidden sensitive information.
\rev{Recently, several studies~\cite{zhu2024adaptive,zhao2023fair,moyer2018invariant,creager2019flexibly} utilized mutual information~(MI) to reduce the dependencies between sensitive and pure preference representations.}
While most of these works assume that users' sensitive attributes are readily available during model training, a subset of studied~\cite{dai2021say,zhang2023fairlisa} has highlighted the limitations of this assumption~\cite{zhang2023fairlisa} and explored achieving fairness with only limited sensitive information.

However, in practice, such sensitive information can be entirely unavailable.
For example, music streaming services like Spotify~\cite{jacobson2016music} usually only gather data that directly enhances user experience, such as musical preference, rather than personal attributes like gender, age or race.
Additionally, on completely anonymous platforms, such as Hush\footnote{\href{https://hush.ac/}{https://hush.ac/}}, no personal information that could identify users in real life is collected at all.
%
%
Unfortunatly, training fair recommendation models without any guidance of sensitive attributes poses significant challenges.
Firstly,
\textbf{uncovering underlying sensitive patterns from complex user behaviors is difficult without labeled sensitive attributes.}
In their absence, such patterns must be inferred indirectly through correlations between behaviors and the hidden sensitive structure in the data.
For example, recent studies~\cite{zeng2024fair} leverage training gradients as proxies for these correlations, assuming that larger gradients reflect more stereotypical and potentially unfair patterns.
However, due to the inherently diverse and multifaceted nature of user behaviors, these simple correlations tend to be supurious and unreliable.
%
Secondly,
\textbf{estimated sensitive patterns often carry noise and inaccuracies, which can hinder the fairness training process.}
As the true sensitive patterns are hard to uncover, the approximations may misrepresent the underlying structure, causing models to learn from suboptimal signals and ultimately degrading fairness performance.

Recently, large language models~(LLMs) trained on extensive data have demonstrated impressive world knowledge and excelled in various tasks~\cite{luo2023reasoning,robinson2022leveraging,menon2022visual}.
Since their training corpus often contains user profiles and preferences, they could inherently enable an understanding of sensitive information beyond the implicit correlations within the data.
Therefore, we propose an \underline{LLM}-enhanced framework for \underline{F}air recommendation with\underline{O}ut \underline{S}ensitive \underline{A}ttributes, named \model.
Specifically, we first propose a \textit{Multi-Persona Sensitive Information Inference} module to identify underlying sensitive patterns.
Given the complex nature of user behaviors, we employ multiple LLM-powered annotators with distinct personas to mirror diverse aspects of human understanding and infer sensitive attributes.
These multifaceted inferences enable a comprehensive analysis of complex user behaviors, improving inference precision while mitigating individual biases.
A meta summarizer further highlights the rationales behind the inference, enhancing sensitive information discovery.
%
To address potential inaccuracies in the inference, we devise a \textit{Confusion-Aware Sensitive Representation Learning} module that refines robust sensitive representations from the inference.
This module captures each annotator's mislabeling confusion and their collective consensus to correct sensitive distributions under noisy inferences. 
Inference rationales are further incorporated to enhance the sensitive representation learning.
Finally, the model is optimized in a two-stage manner by the formulated MI optimization objective.
The contributions of this work are highlighted as follows:
\begin{itemize}[leftmargin=9mm]
	\item We propose an LLM-enhanced framework to effectively improve recommendation fairness without sensitive attributes, tackling a practical scenario where no personal profile is disclosed.
	\item We devise a Multi-Persona Sensitive Information Inference module to uncover latent sensitive patterns without explicit labels. This module employs multi-persona LLMs to mimic diverse human perceptions, enabling a comprehensive and unbiased inference of sensitive information.
	\item We introduce a Confusion-Aware Sensitive Representation Learning module to develop robust sensitive representations under noisy inferences by considering annotators' mislabeling confusion, their collective consensus, and incorporating fine-grained inference rationales.
	\item We conduct extensive experiments on two public real-world datasets to validate \model's superiority in improving recommendation fairness without sensitive attributes.
\end{itemize}

\vspace{-2mm}
\section{Preliminaries}
\vspace{-2mm}

\subsection{Problem Definition}
\vspace{-2mm}
We denote the user set and item set as $\mathcal{U}$ ($|\mathcal{U}|=M$) and $\mathcal{V}$ ($|\mathcal{V}|=N$), respectively.
Let $\mathcal{R} \in \{0,1\}^{|\mathcal{U}| \times |\mathcal{V}|}$ represent the user-item interaction matrix, with entries denoted by $r_{uv}$.
In this study, we focus on the implicit feedback scenario, where $r_{uv} = 1$ indicates that the user $u \in \mathcal{U}$ has an observed interaction with item $v \in \mathcal{V}$, and $r_{uv} = 0$ otherwise.
We denote the set of historical items that user $u$ has interacted with as $h(u)$.
The core of modern RSs is to learn representative embeddings for users and items, defined as: 
$\rep{U}, \rep{V} = \mathcal{E}(\mathcal{U}, \mathcal{V}, \mathcal{R}) = [\rep{u}_1, \rep{u}_2, \ldots, \rep{u}_M ], [ \rep{v}_1, \rep{v}_2, \ldots, \rep{v}_N ]$,
where $\mathcal{E}(\cdot)$ is a collborative encoder.
Subsequently, the preference prediction $\hat{r}_{uv}$ of a user $u$ over an item $v$ can be obtained based on the learned embeddings (\eg $\hat{r}_{uv} = \rep{u}^{\mathrm{T}}\rep{v}$ in matrix factorization models).
Frequently used notations are summarized in \cref{tab:notation}.

When addressing fairness in recommendations, the goal is to ensure that \textit{the items recommended based on the user embedding $\rep{u}$ are independent of the user's sensitive attributes.}
%
%
Specifically, our study aims to improve recommendation fairness without relying on any sensitive attributes.
Formally, let $a_u$ represent the sensitive attribute of user $u$, taking values from the sensitive attribute set $\mathcal{A}$, \ie $a_u \in \mathcal{A}$. We state our problem as follows:
\begin{prbl}[Fair Recommendation]
	Given the user set $\mathcal{U}$, item set $\mathcal{V}$ and the interaction matrix $\mathcal{R}$, train a fair recommender $\mathcal{F}$ without knowing the sensitive attribute $a_u$ for any user $\forall u \in \mathcal{U}$. The recommender $\mathcal{F}$ should deliver \rev{
 fair recommendations which are independent of the user's unobservable sensitive attribute $a_u$.
    }
\end{prbl}

\begin{figure*}[t]
	\centering
	\includegraphics[width=\textwidth]{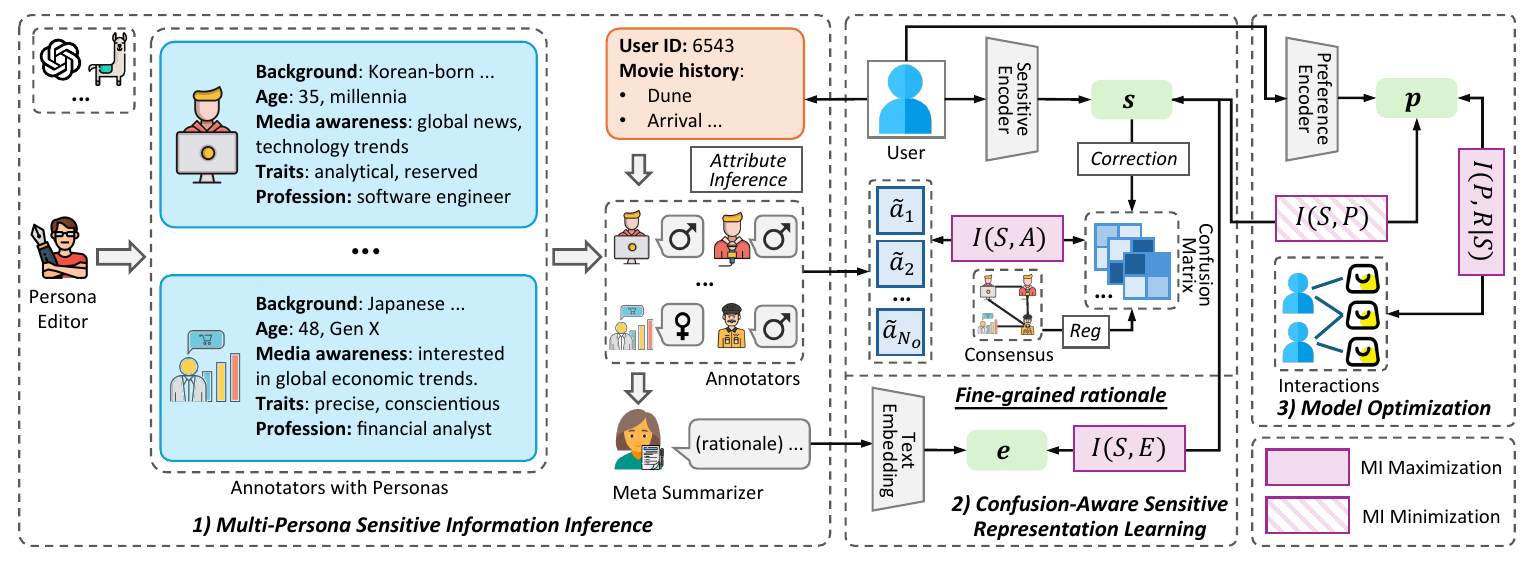}
	\vspace{-7mm}
	\caption{
	The overview of the \model framework. 1) LLM-powered agents collaboratively infer users' sensitive information through diversified roles. 2) Inference results and rationales are incorporated, considering mislabeling confusion and collective consensus to develop robust sensitive representations. 3) Fairness is achieved by optimizing  the formulated MI objective.
	}
	\vspace{-3mm}
	\label{fig:framework}
  \end{figure*}

\vspace{-1mm}
\subsection{Mutual Information Formulation}
\vspace{-2mm}
Inspired by the success of recent MI-based fairness studies~\cite{zhao2023fair,zhu2024adaptive,song2019learning,creager2019flexibly}, we propose to achieve recommendation fairness by formulating it as an MI optimization problem, which allows for precise control over the mutual influence of information.
Following~\cite{zhao2023fair}, for each user $u \in \mathcal{U}$, the original user embedding $\rep{u}$ is first disentangled into the \textit{sensitive-aware} embedding $\rep{s}_u$ and the \textit{sensitive-blind} embedding $\rep{p}_u$.
The sensitive-aware embedding $\rep{s}_u$ contains sufficient sensitive information that could lead to unfair recommendations, while the sensitive-blind embedding $\rep{p}_u$ is designed to be independent of sensitive information, making it suitable for fair recommendations.
For notational convenience, we denote $S$ as the stochastic variable that takes values $\{\rep{s}_u|u \in \mathcal{U}\}$, and $P$ as the stochastic variable that takes values $\{\rep{p}_u|u \in \mathcal{U}\}$.
Moreover, we let $A$ and $R$ be the stochastic variables representing sensitive attributes and user-item interactions, respectively.
Formally, the MI optimization problem proposed in this work is given by:
\begin{equation}
\label{eq:mi_objective}
	\min_{S,P}{-I(S; A) + I(S;P) - I(P; R|S)}. 
\end{equation}
Specifically, maximizing $I(S;A)$ encourages the sensitive-aware embeddings to retain as much sensitive information as possible.
Minimizing $I(S;P)$ ensures that the sensitive-aware embeddings contain information distinguishable from that in sensitive-blind embeddings.
Moreover, maximizing the conditional MI $I(P;R|S)$ promotes the learning of collaborative knowledge in sensitive-blind embeddings while avoiding contamination by sensitive information.
In the following sections, We will elaborate on how to effectively estimate these MI terms, particularly $I(S;A)$,  given the absence of sensitive attributes during training in our problem setting.

\vspace{-2mm}
\section{Methodology}
\vspace{-2mm}
Our proposed framework illustrated in \cref{fig:framework} aims to uncover latent sensitive patterns and leverage them to facilitate fairness training when explicit sensitive attributes are unavailable.
Given the complexity of user behaviors, the \textit{Multi-Persona Sensitive Information Inference} (\cref{sec:llm}) module employs LLM-powered agents to mimic diverse human perceptions, providing a comprehensive analysis of users' sensitive information.
Since these potentially flawed inferences may lead to suboptimal sensitive distributions, the \textit{Confusion-Aware Sensitive Representation Learning} module (\cref{sec:sensitivity}) further incorporates inference rationales and models the agents' confusion and consensus, similar to human reasoning, to develop robust sensitive representations for fairness training.
The fairness is ultimately achieved through the model trained with the formulated MI objectives (\cref{sec:mi}).
The dicussion on our framework's applicability is provided in \cref{app:applicability}.

\vspace{-1mm}
\subsection{Multi-Persona Sensitive Information Inference}
\label{sec:llm}
\vspace{-2mm}
\label{sec:llm}
Given the absence of sensitive information during training, we leverage LLMs with their reasoning abilities to infer the missing sensitive information.
We first create a \textit{persona editor} agent to characterize several \textit{annotator} agents, each with distinct personas.
These annotators work concurrently to infer users' sensitive attributes, providing a comprehensive analysis of user profiles.
To further enhance sensitive pattern discovery by deepening the understanding of these inferences, a \textit{meta summarizer} agent is employed to distill, highlight and provide detailed rationales for the inferences.

{\textbf{Persona Editor.}}
Individual perceptions of subjective inferences vary significantly due to diverse backgrounds and experiences. 
For instance, when identifying gender from ambiguous user profiles, an anthropologist might prioritize cultural cues, while a sociologist may focus on societal roles.
This diversity in perspectives yields varied interpretations and conclusions about the same subject.
A broad spectrum of viewpoints can fully unveil the subject and counteracts the bias that any single viewpoint might introduce.
To harness this diversity, we design the persona editor, an agent who generates a variety of personas, each reflecting unique background knowledge and interpretative biases.
These diverse personas enable more comprehensive and unbiased inferences of sensitive information.

Specifically, we employ a prompting method that enables LLMs to act as a persona editor, generating a diverse set of personas covering various background aspects: $\mathcal{O} = \left\{o_1, \ldots, o_{N_o}\right\} =
	\texttt{LLM}\left( T^{pe}(N_{o}) \right),$
where $\mathcal{O} = \left\{o_1, o_2, \ldots, o_{N_o}\right\}$ is the generated personas, $N_{o}$ is the number of personas, and $T^{pe}(\cdot)$ is the prompt template for persona editor which can be found in \cref{app:persona_editor}.

{\textbf{Annotators.}}
Using the generated personas, we develop a set of annotator agents tasked with inferring sensitive attributes by analyzing users' historical behaviors.
Formally, given an annotator with persona $o_i \in \mathcal{O}$ and the historical behaviors $h(u)$ of user $u \in \mathcal{U}$, the sensitivite attribute inference is given by $\tilde{c}^i_u = \texttt{LLM}(T^{inf}(o_i, h(u))),$
%
where $T^{inf}(\cdot)$ is the prompt template for each annotator which can be found in \cref{app:annotator_with_persona}.
Typically, the response $\tilde{c}_u^i$ from an annotator includes their inferred attribute and a brief reason for their judgment.
To standardize these outputs, we employ a verbalizer to convert the textual response $\tilde{c}_u^i$ into a categorial label $\tilde{a}_u^i \in \left\{ 0,1,\ldots, |\mathcal{A}| - 1 \right\}$ within the attribute space.

{\textbf{Meta Summarizer.}}
Beyond the sensitive attributes inferred by the annotators, the fine-grained rationales behind each annotator's decision-making process are crucial for understanding the users' sensitive information.
To thoroughly capture and distill these detailed rationales, we introduce a meta summarizer agent.
This agent analyzes the inference processes of each annotator, aggregating their insights to highlight key aspects of the decision-making process.
The meta summarizer provides a coherent summary that reveals the rationale behind the determination of a user's sensitive attributes.
\ir{Moreover, these human-readable summaries also enhance the interpretability of our framework.}

Specifically, given a user $u \in \mathcal{U}$ with the historical interactions $h(u)$ and the sensitive attribute inference $\tilde{c}_u^i$ from each annotator $\forall o_i \in \mathcal{O} $, the meta summarizer performs as $e_u = \texttt{LLM}\left(T^{sum}\left(h(u), \tilde{c}_u^1, \tilde{c}_u^2, \ldots, \tilde{c}_u^{N_o}\right)\right),$
where $T^{sum}(\cdot)$ is the prompt for the meta summarizer presented in \cref{app:meta_summarizer}.
Annotator and meta summarizer's response examples are in \cref{app:response}.


\vspace{-1mm}
\subsection{Confusion-Aware Sensitive Representation Learning}
\vspace{-2mm}
\label{sec:sensitivity}
Due to the hallucination issue~\cite{zhang2023siren} with LLMs, the inferred sensitive information may be potentially inaccurate and  compromise the fairness training.
To counter this, we now introduce how to utilize these inferences to develop robust and effective sensitive-aware embeddings.
Recall that we have obtained user and item embeddings $\rep{u}, \rep{v}$ from the collaborative encoder $\mathcal{E}(\cdot)$.
For each user embedding $\rep{u}$, we employ a sensitive encoder $\mathcal{S}(\cdot)$ to transform it into the sensitive-aware embedding $\rep{s}_u$ as:
$
	\rep{s}_u = \mathcal{S}\left( \rep{u} \right), u \in \mathcal{U},
$
where $\mathcal{S}(\cdot)$ is implemented as a multi-layer perceptron (MLP).
As shown in \cref{eq:mi_objective}, we will learn the sensitive-aware embeddings by maximizing the MI $I(S;A)$.

{\textbf{Confusion Modeling.}}
While diverse annotator personas offer rich insights into users' sensitive attributes,
they can also cause variations in sensitive attribute inferences, potentially leading to inconsistencies in the sensitive representation learning process.
Therefore, we design a learnable confusion matrix for each annotator to model confusion probabilities and correct the MI estimation, ultimately learning robust representations.
In our case, the actual sensitive attribute $a_u$ of the user is missing, making direct computation impractical.
We use the annotators' inferred sensitive attributes to introduce $\widetilde{A}$, an approximation of $A$, which takes values in the inferred sensitive attribute labels.
Specifically, we maximize $I(S; A)$ with the approximation $\widetilde{A}$ by maximizing the lower bound:
\begin{equation}
	\label{eq:sens_bound}
		\max_{\mathcal{S}, \mathcal{C}} \mathbb{E}_{(\rep{s}_u, \tilde{a}_u^i) \sim p(S,\widetilde{A})}
		\left[
		 \log q(\tilde{a}_u^i|\rep{s}_u; \mathcal{C})
		\right],
	\end{equation}
where $\mathcal{C}(\cdot)$ is the classifier used to classify the sensitive label based on the sensitive-aware embedding, also implemented as an MLP. The derivation is provided in \cref{app:a_bound}.
%

To introduce the confusion modeling, we first adopt the common assumption~\cite{dawid1979maximum,rodrigues2018deep} that each inference is independent of the user's profile given the ground-truth sensitive attribute.
%
Based on these assumptions, we formally define a confusion matrix $\rep{F}_i$ for each annotator $i$ ($i=1,2,\ldots,N_o$), which models the probabilities of the user's true sensitive attribute being misclassified as the annotator's inferred sensitive attribute:
\begin{equation}
\label{eq:confusion}
	(\rep{F}_i)_{jk} \triangleq q(\tilde{a}_u^i = k | a_u = j; \rep{F}_i), \forall u \in \mathcal{U}. 
\end{equation}

Using the confusion matrices in \cref{eq:confusion}, we can further decompose \cref{eq:sens_bound} into:
\begin{equation}
\label{eq:mi_coarse_final}
	\max_{\mathcal{S}, \mathcal{C}, \rep{F}} \mathbb{E}_{(\rep{s}_u, \tilde{a}_u^i) \sim p(S,\widetilde{A})}
	\left[
	\log q(\tilde{a}_u^i|a_u; \rep{F}_i)q(a_u|\rep{s}_u;\mathcal{C})
	\right].
\end{equation}
Finally, we derive the loss function that maximizes the objective in \cref{eq:mi_coarse_final} to learn robust sensitive-aware embeddings:
\begin{equation}
	\mathcal{L}_{cls} = \sum_{u \in \mathcal{U}} \sum_{i=1}^{N_o} 
	\left[
	- \texttt{OneHot}(\tilde{a}_u^i) \cdot \log
	\left(
	\rep{F}_i \cdot \mathcal{C}(\mathcal{S}(\rep{u}))
	\right)
	\right],
\end{equation}

\textbf{Consensus Regularization.}
Individuals with similar personalities often exhibit comparable cognitive patterns.
Based on this premise, we assume that annotators with similar personas are likely to experience similar levels of confusion.
To foster consensus among annotators, we introduce a concise similarity regularization term to align the confusion matrices of annotators with similar personas.
Specifically, we first define the annotator similarity for each annotator persona $o_i \in \mathcal{O}$ as:
\begin{equation}
\label{eq:similarity}
	\Gamma(o_i) = \left\{ o_j| 
	d(o_i, o_j)  \leq d(o_i, \bar{o}_{i,K}) 
	\right\}, \quad
	d(o_i, o_j) = \| \texttt{TxtEmb}(o_i) - \texttt{TxtEmb}(o_j) \|_2, 
\end{equation}
where $\texttt{TxtEmb}(\cdot)$ is a text embedding model, $\bar{o}_{i,K}$ denotes the $K$-th nearest persona to $o_i$, with $K$ being a hyperparameter.
The regularization term based on the annotator similarity is defined as:
\begin{equation}
	\mathcal{L}_{sim} = \sum_{o_i \in \mathcal{O}} \sum_{o_j \in \Gamma(o_i)}
	\| \rep{F}_i - \rep{F}_j \|_2.
\end{equation}

\textbf{Fine-Grained Rationale Incorporation.}
The decision-making rationales of annotator agents are crucial for sensitive pattern discovery, as they provide key insights into sensitive attribute inference.
However, the direct incorporation of text-based rationales poses challenges due to their heterogeneous nature.
To address this, we first utilize a text embedding model to convert the textual rationales into low-dimensional, semantic embeddings:
$
	\rep{e}_u = \texttt{TxtEmb}(e_u),
$
where $e_u$ is the rationale concluded by the meta summarizer, as described in \cref{sec:llm}.
These encoded representations serve as proxies for the fine-grained rationales underlying sensitive information inference.
We propose maximizing the MI between sensitive and rationale representations, thereby optimizing the integration of rationale knowledge into the sensitive-aware representation.
Formally, let $E$ be the variable taking values $\{\rep{e}_u|u \in \mathcal{U}\}$. 
For effective optimization~\cite{oord2018representation}, we employ the lower bound of $I(S;E)$ and apply contrastive learning methods to develop sensitive-aware embeddings. The lower bound is given by:
\begin{equation}
	I(S;E) \geq \mathbb{E}\left[\log \frac{\texttt{DR}(\rep{s}_u, \rep{e}_u)}{\sum_{\rep{s} \sim p(S)} \texttt{DR}(\rep{s}, \rep{e}_u)} \right],
\end{equation}
where $\texttt{DR}(\cdot)$ is the density ratio implemented via inner product.
This approach allows us to maximizie $I(S;E)$ by optimizing the sensitive-aware embedding $\rep{s}_u$ using the following contrastive loss:
\begin{equation}
	\mathcal{L}_{fine} = - \sum_{u \in \mathcal{U}}
	\log \frac{\exp(\rep{s}_u^\mathrm{T}\rep{e}_u)}{\sum_{j \in \mathcal{U}} \exp( \rep{s}_j^\mathrm{T}\rep{e}_u)}.
\end{equation}

After defining the individual loss functions, we construct a composite loss function to simultaneously learn the sensitive-aware embeddings as:
\begin{equation}
\label{eq:sens_overall}
	\mathcal{L}_{sen} = \mathcal{L}_{cls} + \mathcal{L}_{bpr} + \lambda_{sim} \mathcal{L}_{sim} + \lambda_{fine}\mathcal{L}_{fine},
\end{equation}
where $\lambda_{fine}, \lambda_{sim}$ are two hyperparameters, and $\mathcal{L}_{bpr}$ is the Bayesian personalized ranking~(BPR)~\cite{rendle2012bpr} loss, which trains the collaborative encoder as:
\begin{equation}
\label{eq:bpr}
	\mathcal{L}_{bpr} = - \sum_{u \in \mathcal{U}} \sum_{v \in h(u)}
	\log\left(
	\sigma\left(
	\rep{u}^{\mathrm{T}}\rep{v} - 
	\rep{u}^{\mathrm{T}}\rep{v}_{-}
	\right)
	\right),
\end{equation}
where $\sigma$ is the activation function and $v_{-}$ is a random negative item for each positive item $v \in h(u)$.

\vspace{-1mm}
\subsection{Model Optimization}
\vspace{-2mm}
\label{sec:mi}
As outlined in \cref{eq:mi_objective}, the model optimization objective comprises MI terms including $I(S;A)$, $I(S;P)$ and $I(P;R|S)$.
\cref{sec:sensitivity} has introduced a novel approach for optimizing $I(S;A)$.
This section will describe how to estimate the remaining MI terms and present the overall training process.

For each user embedding $\rep{u}$, we apply a preference encoder $\mathcal{P}(\cdot)$ implemented as an MLP to generate the sensitive-blind embedding
$
	\rep{p}_u = \mathcal{P}(\rep{u}), u \in \mathcal{U}.
$
%
Then we optimize the following MI terms.

\textbf{Minimizing $I(S;P)$.}
Following~\cite{zhao2023fair}, we employ the contrastive log-ratio upper bound~\cite{cheng2020club} of MI to derive the following loss function:
\begin{equation}
	\label{eq:loss_ub}
	\mathcal{L}_{ub} = \sum_{u \in \mathcal{U}} \left[
	\log \mathcal{B}(\rep{p}_u) - \frac{1}{M} \sum_{u^\prime \in \mathcal{U}} \log \mathcal{B}(\rep{p}_{u^\prime})
	\right],
\end{equation}
where $\mathcal{B}(\cdot)$ is a neural network implemented as an MLP.
The derivation can be found in \cref{app:mi_ub}.
%

\textbf{Maximizing $I(P;R|S)$.}
We first use a pre-trained collaborative filtering (CF) model to obtain the embedding $\rep{r}_u$, representing variable $R$ of interactions.
We apply conditional InfoNCE~\cite{gupta2021controllable} to maximize the lower bound of conditional MI and derive the following loss function:
\begin{equation}
\label{eq:lb}
	\mathcal{L}_{lb} = - \sum_{u \in \mathcal{U}} \left[
	\log \frac{\exp\left({\langle\rep{r}_u, \rep{p}_u + \alpha \rep{s}_u\rangle}\right)}{\frac{1}{M} \sum_{u^{\prime} \in \mathcal{U}} \exp\left({\langle\rep{r}_{u^\prime}, \rep{p}_u + \alpha \rep{s}_u\rangle}\right) }
	\right],
\end{equation}
where $\langle\cdot\rangle$ denotes cosine similarity and $\alpha$ is a parameter. The derivations can be found in \cref{app:mi_lb}.

Finally, we employ a two-stage approach to train the \model framework.
In the \textit{first} stage, we use the loss function $\mathcal{L}_{sen}$ in \cref{eq:sens_overall} to maximize $I(S;A)$.
Once the sensitive-aware embedding $\rep{s}_u$ captures sufficient sensitive information, we proceed to the \textit{second} stage, using the following loss function to jointly optimize $I(S;P)$ and $I(P;R|S)$ in an information bottleneck manner~\cite{shwartz2017opening}:
\begin{equation}
\label{eq:bound}
	\mathcal{L}_b = \mathcal{L}_{bpr} + \lambda_{ub}\mathcal{L}_{ub} + \lambda_{lb}\mathcal{L}_{lb},
\end{equation}
where $\mathcal{L}_{bpr}$, defined in \cref{eq:bpr}, helps in learning representative user embeddings, and $\lambda_{ub}$ and $\lambda_{lb}$ are hyperparameters.
Upon optimal training of \model, the learned sensitive-blind embedding $\rep{p}_u$ is used to recommend items to users to achieve fairness.


\vspace{-1mm}
\section{Experiments}
\vspace{-1mm}
\subsection{Experimental Settings}
\vspace{-2mm}
{\textbf{Datasets.}}
Following previous research~\cite{zhao2023fair,chen2023improving,wu2021learning}, we conduct experiments on two public datasets: MovieLens-1M(\textbf{\ml})~\cite{harper2015movielens} and LastFM-360K(\textbf{\lastfm})~\cite{celma2009music}.
Following~\cite{chen2023improving,zhao2023fair}, we consider gender as the sensitive attribute. 
Analysis on other sensitive attributes is also provided in \cref{app:occupation}.
Note that, since sensitive attributes are unavailable during training, they are only used in the test set for evaluation purposes.
Detailed dataset statistics and preprocessing steps are provided in \cref{app:dataset}.

\begin{table*}[t]
  \centering
  \renewcommand\arraystretch{1.1}
   \tabcolsep 0.07in
  \caption{The overall fairness (AUC) and recommendation accuracy (N@20 and R@20) performance.}
    \scalebox{0.7}{
      \begin{tabular}{c|>{\columncolor[rgb]{0.898,0.898,0.898}}c|cc|>{\columncolor[rgb]{0.898,0.898,0.898}}c|cc|>{\columncolor[rgb]{0.898,0.898,0.898}}c|cc|>{\columncolor[rgb]{0.898,0.898,0.898}}c|cc}
        \toprule
        \textit{Dataset}& \multicolumn{6}{c|}{\textbf{\ml}} & \multicolumn{6}{c}{\textbf{\lastfm}} \\
   \midrule
        \textit{Backbone} & \multicolumn{3}{c|}{MF} & \multicolumn{3}{c|}{GCN} & \multicolumn{3}{c|}{MF} & \multicolumn{3}{c}{GCN} \\
   \midrule
        \textit{Metrics} & AUC($\downarrow$)  & R($\uparrow$) & N($\uparrow$) & AUC($\downarrow$)  & R($\uparrow$) & N($\uparrow$) & AUC($\downarrow$)  & R($\uparrow$) & N($\uparrow$) & AUC($\downarrow$)  & R($\uparrow$) & N($\uparrow$) \\
        \midrule
   NoFair & 0.7163 & 0.2297 & 0.2310 & 0.7125 & 0.2363 & 0.2376 & 0.6726 & 0.2274 & 0.1710 & 0.6822 & 0.2500 & 0.1920 \\
   \midrule
   ARL  & 0.7166 & 0.2289 & 0.2305 & 0.7212 & 0.2418 & 0.2426 & 0.6747 & 0.2264 & 0.1707 & 0.6854 & 0.2506 & 0.1917 \\
   FKD  & 0.7098 & 0.2276 & 0.2286 & 0.7208 & 0.2365 & 0.2398 & 0.6694 & 0.2213 & 0.1643 & 0.6732 & 0.2416 & 0.1846 \\
   EIIL & 0.7230 & 0.2398 & 0.2413 & 0.7091 & 0.2238 & 0.2292 & 0.6739 & 0.2289 & 0.1722 & 0.6889 & 0.2456 & 0.1888 \\
   \midrule
   ComFair & 0.7045 & 0.2140 & 0.2186 & 0.6823 & 0.2134 & 0.2195 & 0.6687 & 0.2093 & 0.1573 & 0.6824 & 0.2205 & 0.1666 \\
   FairGo & 0.7105 & 0.2227 & 0.2259 & 0.6774 & 0.2155 & 0.2219 & 0.6727 & 0.2130 & 0.1606 & 0.6814 & 0.2200 & 0.1658 \\
   FDA  & 0.7234 & 0.2229 & 0.2184 & 0.7298 & 0.2246 & 0.2229 & 0.6851 & 0.2227 & 0.1696 & 0.7026 & 0.2158 & 0.1618 \\
   FairMI & 0.7207 & 0.2321 & 0.2346 & 0.7116 & 0.2205 & 0.2268 & 0.6722 & 0.2212 & 0.1666 & 0.6845 & 0.2421 & 0.1857 \\
   \midrule
   FairGNN & 0.6889 & 0.2052 & 0.2047 & 0.7003 & 0.2159 & 0.2175 & 0.6448 & 0.2052 & 0.1533 & 0.6668 & 0.2218 & 0.1695 \\
   FairLISA & 0.7066 & 0.1867 & 0.1900 & 0.6990 & 0.1875 & 0.1934 & 0.6424 & 0.1702 & 0.1228 & 0.6629 & 0.1890 & 0.1399 \\
   AFSR & 0.6490 & 0.1438 & 0.1572 & 0.7018 & 0.2295 & 0.2351 & 0.6484 & 0.1452 & 0.1062 & 0.6823 & 0.2404 & 0.1833 \\
   \midrule
   LLMFOSA & \textbf{0.5974} & 0.2267 & 0.2296 & \textbf{0.6253} & 0.2205 & 0.2259 & \textbf{0.6305} & 0.2025 & 0.1536 & \textbf{0.6551} & 0.2330 & 0.1798 \\
   \bottomrule
   \end{tabular}}%
  \label{tab:overall}%
  \vspace{-4mm}
\end{table*}%

{\textbf{Baselines.}}
we compare \model against several baselines, including
\textit{1) general fairness methods without demographics:}
ARL~\cite{lahoti2020fairness}, FKD~\cite{chai2022fairness}, EIIL~\cite{creager2021environment};
\textit{2) recommendation fairness methods with full sensitive attributes:}
ComFair~\cite{bose2019compositional}, FairGo~\cite{wu2021learning}, FDA~\cite{chen2023improving}, FairMI~\cite{zhao2023fair};
\textit{3) recommendation fairness methods with limited or no sensitive attributes:}
FairGNN~\cite{dai2021say}, FairLISA~\cite{zhang2023fairlisa}, AFSR~\cite{zeng2024fair}.
The baseline descriptions are provided in \cref{app:baseline}.
\ir{Notably, the fairness methods with full sensitive attributes generally rely on explicit sensitive attributes during training, which are unavailable in our setting.
To ensure a fair evaluation under this constraint, we propose extracting sensitive patterns from user behavior~\cite{lahoti2020fairness}.
Specifically, we apply hierarchical clustering on user embeddings to generate cluster labels as proxies for sensitive attributes.
Implementation details and comparisons with alternative clustering methods are provided in \cref{sec:accuracy}.}
Moreover, we apply \model and the baseline models on two CF backbone models, \ie, MF~\cite{rendle2012bpr} and LightGCN (GCN)~\cite{he2020lightgcn}.

{\textbf{Evaluation Protocols.}}
We evaluate both recommendation accuracy and fairness.
Recommendation accuracy is assessed using two widely used metrics, \ie NDCG (N@20)~\cite{jarvelin2017ir} and Recall (R@20).
For recommendation fairness, following~\cite{zhang2023fairlisa,bose2019compositional,wu2021learning}, we train an attacker to identify sensitive information.
Detailed fairness evaluation methods are provided in \cref{app:fairness_evaluation}.
Moreover, the implementation details of \model are provided in \cref{app:implementation}.

\vspace{-1mm}
\subsection{Overall Performance}
\vspace{-2mm}
\label{sec:overall_performance}
We conducted experiments on the \ml and \lastfm datasets to evaluate the effectiveness of our \model framework in improving recommendation fairness.
The overall performance of \model, compared to baseline models, is presented in \cref{tab:overall}, where the mean results are derived from three random runs. 
The NoFair refers to a naive recommendation backbone without any fairness considerations. 
Particularly, we observed several key findings:

\begin{itemize}[leftmargin=5mm]
	\item \model outperforms all baselines in fairness, confirming its effectiveness even without sensitive information.
\ir{In contrast, baseline methods that infer sensitive attributes relying to the varing degrees on complex user behaviors without any semantic cues, experience a significant decline in fairness, especially when behavior poorly correlates with sensitive attributes.}

\item \ir{FairGNN and FairLISA perform relatively well in fairness by partially relying on sensitive attributes, making them less prone to error from imprecise inferences. However, this reduced dependence compromises recommendation accuracy. Moreover, these methods still require some sensitive data, limiting their practical applicability.
In contrast, AFSR operates without sensitive attributes and achieves promising fairness outcomes. Nevertheless, it suffers from a high accuracy drop, as its approach of smoothing item embeddings can distort the original data semantics.}

\item The naive NoFair method surprisingly shows competitive fairness compared to some baselines that explicitly address fairness.
This is possibly because fairness-aware models alter embeddings based on inferred sensitive labels. If these labels are inaccurate, such adjustments can misfire.

	\item \model shows a modest drop in accuracy compared to some fairness baselines.
This is likely because achieving fairness with relatively precise sensitive information in our framework can compromise more recommendation performance.
Nonetheless, \model achieves a significant fairness boost (up to 13.3\%) in scenarios where sensitive information is completely absent.
\end{itemize}

\vspace{-2mm}
\subsection{Ablation Study}
\vspace{-2mm}
To evaluate the contributions of different components within the \model framework, we conducted a comparative analysis with several variants of \model:
\textbf{1) -inf} does not use any LLM inference results, relying only on random sensitive attributes;
\textbf{2) -per} uses a single annotator's inference as the sensitive attribute;
\textbf{3) -con} removes the confusion matrix in the sensitive representation learning and uses the majority vote as the sensitive attribute;
\textbf{4) -sim} removes the consensus regularization;
\textbf{5) -rat} does not incorporate the fine-grained rationales from the meta summarizer agent.

\begin{figure}[t]
  \centering
  \begin{minipage}{0.45\columnwidth}
    \centering
    \includegraphics[width=0.99\columnwidth]{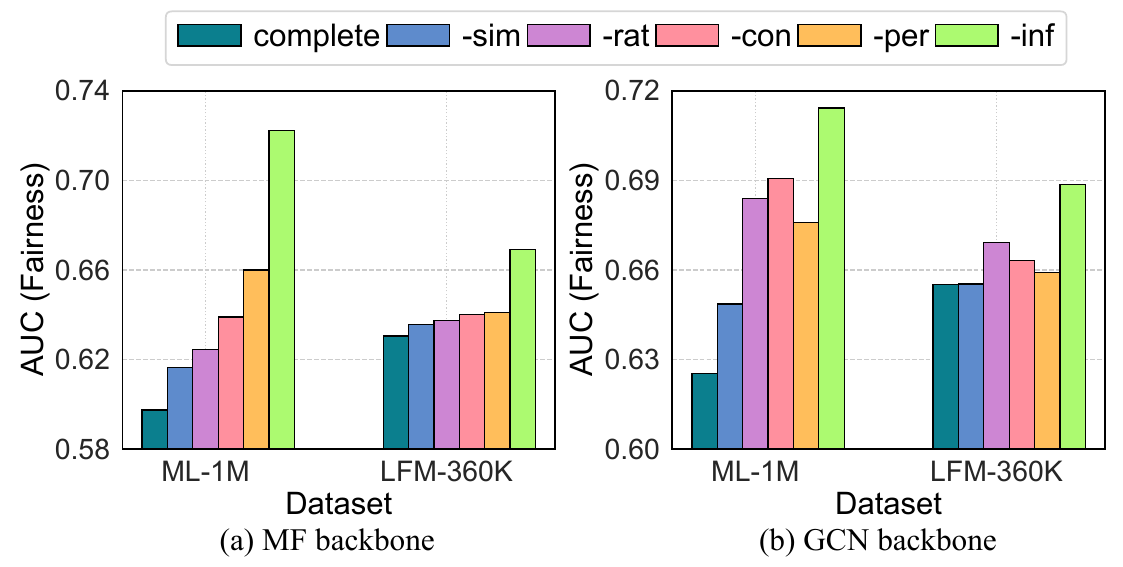}
    \vspace{-7mm}
    \caption{Contributions of different components in \model to the fairness.}
    \label{fig:ablation}
  \end{minipage}
  \hspace{1ex}
  \begin{minipage}{0.52\columnwidth}
    \centering
    \includegraphics[width=0.99\columnwidth]{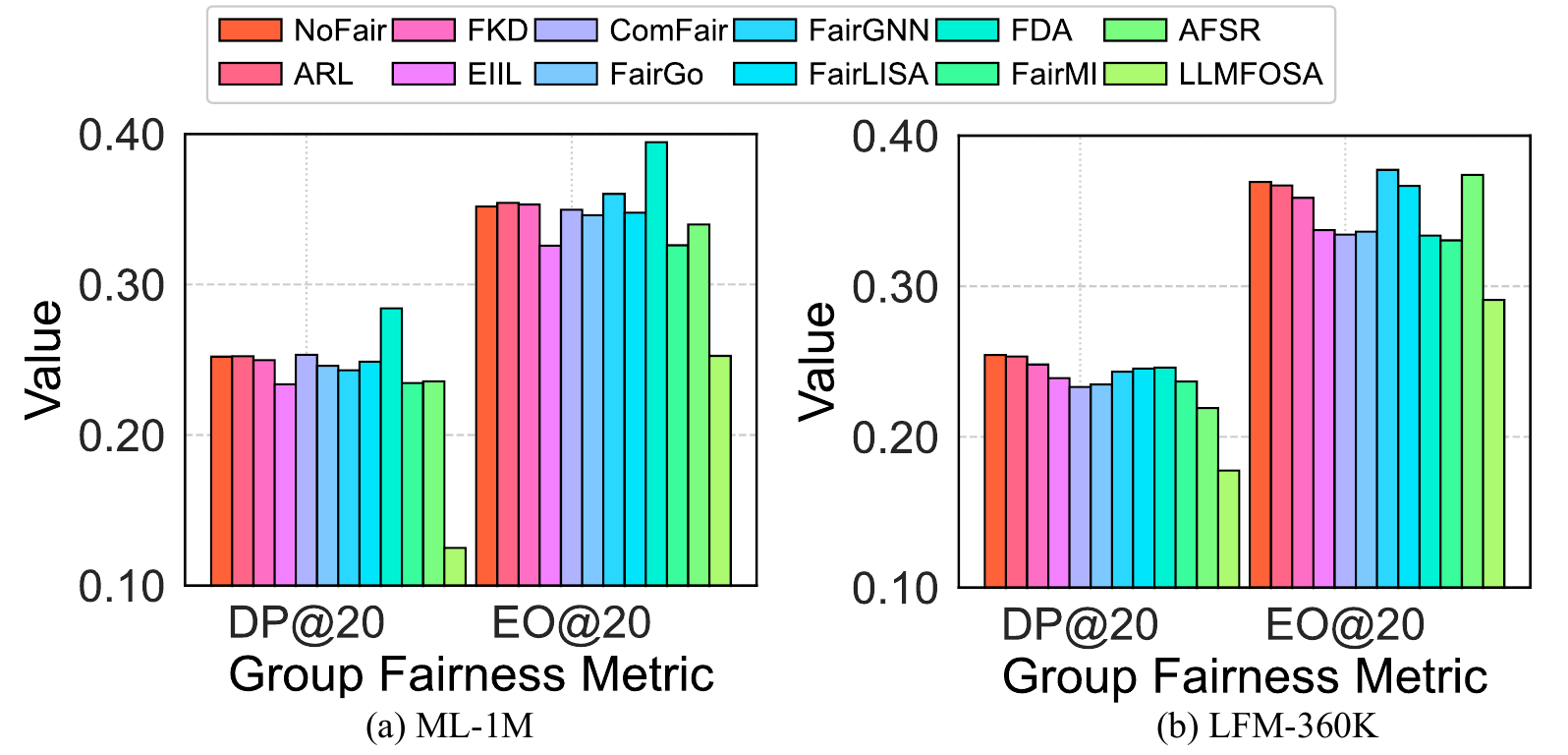}
    \vspace{-8mm}
    \caption{The performance on group fairness metrics.}
    \label{fig:group}
  \end{minipage}
  \vspace{-3mm}
\end{figure}

As shown in \cref{fig:ablation}, \model outperforms all the variants, confirming the effectiveness of its components in improving fairness.
Specifically, omitting LLM inferences entirely or partially (as in \textit{-inf} and \textit{-per}) leads to suboptimal fairness, proving the value of comprehensive sensitive attribute estimation.
The results also show that careful use of LLM-inferred attributes is beneficial.
For instance, the drop in performance for \textit{-con} highlights the importance of modeling annotator confusion for better sensitive representation learning.
Similarly, declines in \textit{-sim} and \textit{-rat} validate the gains from modeling consensus among similar annotators and incorporating fine-grained rationales.
The results of recommendation accuracy are provided in \cref{app:ablation_sensitivity}.

\vspace{-3mm}
\subsection{Group Fairness Evaluation}
\vspace{-2mm}
\ir{To further evaluate whether \model enhances classical group fairness metrics, we assess its performance and compare it against baselines using two widely adopted metrics: Demographic Parity (DP@20)~\cite{dwork2012fairness} and Equal Opportunity~(EO@20)~\cite{hardt2016equality}. Lower values for these metrics indicate better group fairness.
As shown in \cref{fig:group}, \model consistently achieves superior group fairness, outperforming all baselines on both metrics and ensuring fair treatment across different groups.
}

\vspace{-3mm}
\subsection{Attribute Inference Accuracy Analysis}
\vspace{-2mm}
\label{sec:accuracy}
To further evaluate the effectiveness of LLM-inferred attributes in our framework, we compare their accuracy to that of attributes derived from behavior patterns in the dataset. Specifically, we compare their
\begin{wraptable}{r}{0.5\columnwidth}
  \centering
   \tabcolsep 0.09in
   \vspace{-6.5mm}
  \caption{\label{tab:label_acc}Accuracy of sensitive attribute using different strategies.}
    \scalebox{0.78}{
    \begin{tabular}{c|cc|cc}
    \toprule
    \multirow{2}[5]{*}{} & \multicolumn{2}{c|}{\textbf{\ml}} & \multicolumn{2}{c}{\textbf{\lastfm}} \\
 & \multicolumn{1}{c}{Acc} & \multicolumn{1}{c|}{F1} & \multicolumn{1}{c}{Acc} & \multicolumn{1}{c}{F1} \\
    \midrule
        Random & 0.5056 & 0.5953 & 0.4971 & 0.5943 \\
        \midrule
        	K-Means & 0.5669 & 0.6538 & 0.5194 & 0.6243 \\
        	GMM & 0.5853 & 0.7098 & 0.5731 & 0.6869 \\
        	Hierarchical & 0.5818 & 0.6877 & 0.6302 & 0.7570 \\
    \midrule
    	LLM-Single & 0.7192 & 0.7993 & 0.6994 & 0.7913 \\
        LLM-MV & \textbf{0.7644} & \textbf{0.8353} & \textbf{0.7153} & \textbf{0.8047} \\
	\bottomrule
    \end{tabular}}%
    \vspace{-5mm}
\end{wraptable}%
accuracy against three commonly used unsupervised clustering methods: \textbf{K-Means}, Gaussian Mixture Model (\textbf{GMM}) and \textbf{Hierarchical} clustering.
We also include a \textbf{Random} assignment for reference.
For clustering methods, user embeddings from the MF model are used to generate cluster labels, which are then assigned as sensitive attributes. 
For LLM-based attribute inference in LLMFOSA, we employ two strategies: \textbf{LLM-Single}, which assigns a single annotator’s inference as the user’s attribute, and \textbf{LLM-MV}, which aggregates annotations through a majority vote. The accuracy (Acc) and F1 score are used to assess these approaches.

\ir{
As shown in \cref{tab:label_acc}, LLM-inferred attributes outperform both random and the clustering approaches. 
Clustering methods surpassing random assignment, reveals that users’ behaviors can partially expose their sensitive attributes. 
Due to the overall strong performance of hierarchical clustering, we employ it to generate sensitive attributes for fairness baselines that require explicit labels, ensuring fair evaluation. 
Additionally, (LLM-MV) outperforms (LLM-Single), showing the strength of our multi-persona design, which enhances fairness by leveraging collective inference.}

\begin{figure}[t]
    \centering
    \subfigure[\ml, $N_o$]{
        \includegraphics[width=0.23\columnwidth]{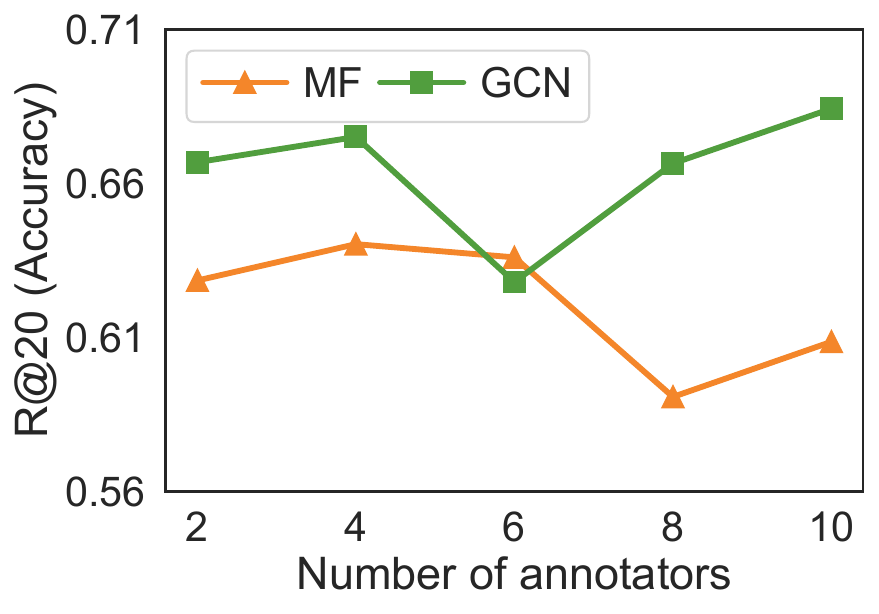}
        	\label{fig:sens_num_ml}
    }\hspace{-1mm}
    \subfigure[\lastfm, $N_o$]{
        \includegraphics[width=0.23\columnwidth]{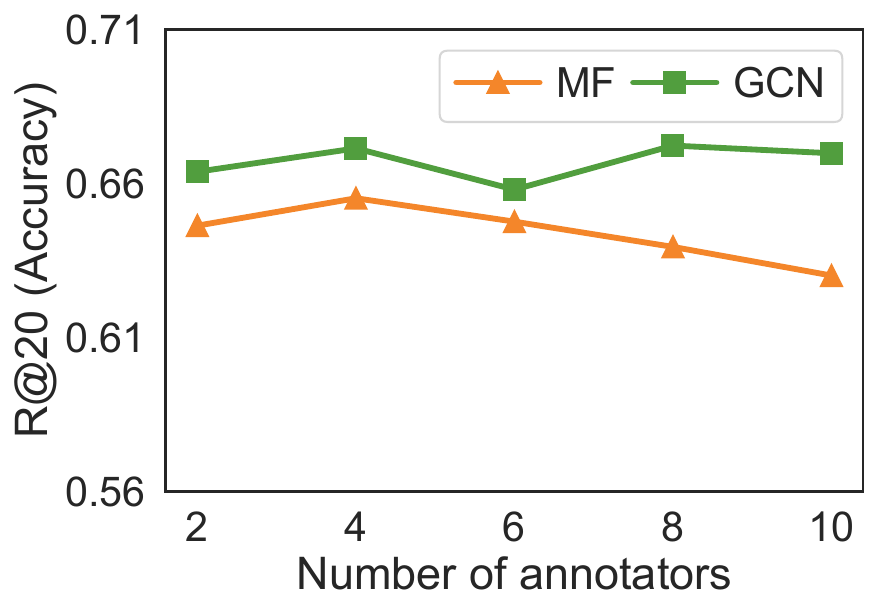}
        	\label{fig:sens_num_lastfm}
    }\hspace{-1mm}
    \subfigure[\ml, $\lambda_{sim}$]{
        \includegraphics[width=0.23\columnwidth]{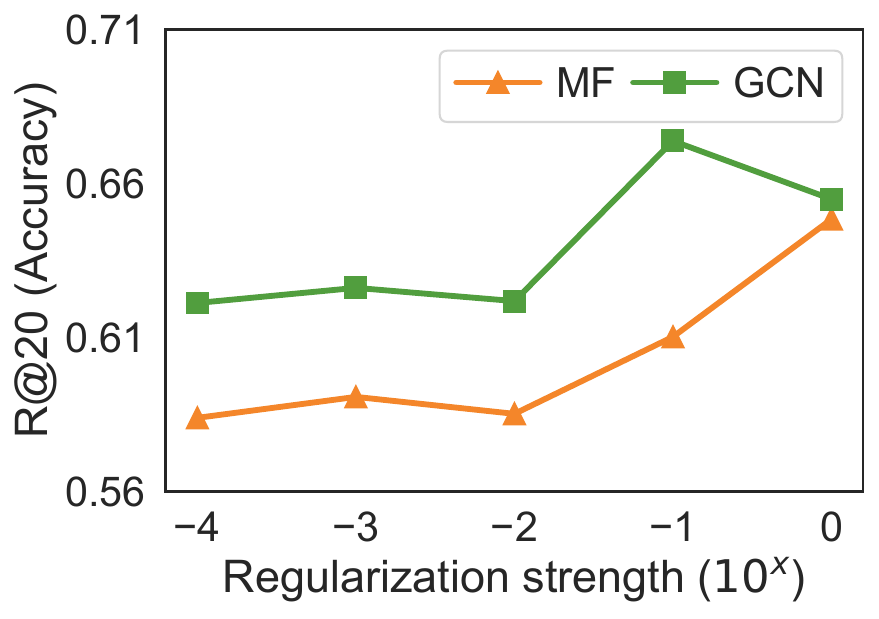}
        	\label{fig:sens_strength_ml}
    }\hspace{-1mm}
    \subfigure[\lastfm, $\lambda_{sim}$]{
        \includegraphics[width=0.23\columnwidth]{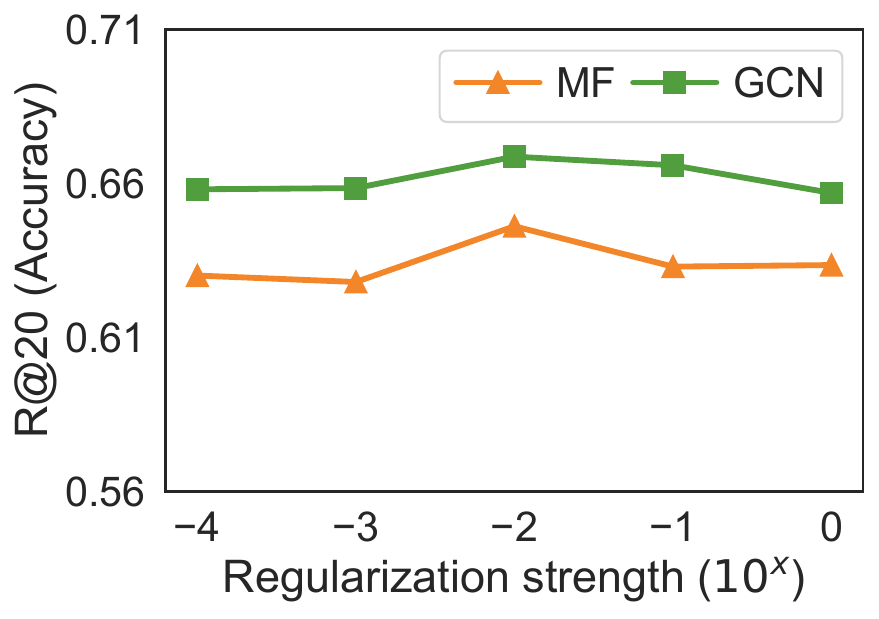}
        	\label{fig:sens_strength_lastfm}
    }
    \vspace{-3mm}
    \caption{Parameter sensitivity on two hyper-parameters. Results of accuracy are in \cref{app:ablation_sensitivity}.}
    \vspace{-5mm}
    \label{fig:sens}
\end{figure}

\vspace{-1mm}
\subsection{Parameter Sensitivity Analysis}
\vspace{-1mm}

\textbf{Number of annotator agents $N_{o}$.}
\ir{As shown in \cref{fig:sens_num_ml,fig:sens_num_lastfm}, fairness initially decreasing and then increasing when $N_o$ exceeds 4.
Specifically, when $N_o$ increases from 2 to 4, noise outweighs sensitive knowledge gains, resulting in a temporary fairness decrease.
As $N_o$ continues to increase, accumulated sensitive knowledge surpasses noise, improving fairness until it eventually declines when knowledge gains diminish while noise continues growing.
Particularly, for LFM-360K with MF as the backbone, this turning point is possibly beyond $N_o=10$.}
Since increasing $N_o$ raises computational costs due to LLM inference overhead, finding an optimal balance between improved fairness and computational efficiency is crucial for practical use.

\textbf{Strength of the similarity-guided regularization $\lambda_{sim}$.}
\cref{fig:sens_strength_ml,fig:sens_strength_lastfm} indicate that stronger regularization tends to decrease fairness in both datasets.
This suggests that overly strong regularization leads to indistinguishable confusion matrices among annotators, potentially reducing the diversity of their decisions and thus impacting fairness negatively.
Interestingly, the \lastfm dataset shows a smaller decrease in fairness compared to the \ml dataset at high $\lambda_{sim}$ values. This may be due to a greater consensus among annotators in the \lastfm dataset, allowing it to maintain better fairness even as the confusion level approaches convergence.

\vspace{-1mm}
\section{Related Works}
\vspace{-1mm}
\textbf{Recommendation Fairness.}
Prior studies~\cite{yao2017beyond,aghaei2019learning} improved fairness by incorporating group fairness metrics as regularizers across user groups.
Another line of research~\cite{bose2019compositional,zhang2023fairlisa,wu2021learning} formulated the problem as a min-max game, using filters and discriminators to eliminate and identify sensitive information.
Recently, some works~\cite{zhu2024adaptive,zhao2023fair,moyer2018invariant,creager2019flexibly} also investigated fairness from an MI perspective. 
%
As sensitive information is not always easily collectible in practice, a few works have addressed fairness with limited~\cite{dai2021say,zhang2023fairlisa} or no~\cite{zeng2024fair} sensitive attributes.
However, these works either still rely on available sensitive information or unreliable correlations between behaviors and sensitive patterns.
In contrast, this work improves fairness without sensitive attributes by leveraging external knowledge from LLMs.

\textbf{LLMs for Recommendation.}
The rapid progress of LLMs has sparked interest in applying them to RSs.
Some works~\cite{wei2024llmrec,ren2024representation} treated LLMs as knowledge bases, retrieving useful information to support conventional recommendation models.
To better leverage training data, numerous works~\cite{bao2023tallrec,lin2024rella,wang2024can,liao2024llara,zhu2024collaborative} have fine-tuned LLMs for recommendation tasks, though this increases computational overhead.
\rev{
Instead of using LLMs as the core engine, we leverage LLM's reasoning power to develop agents that infer users' sensitive information, thereby enhancing recommendation fairness.}

\textbf{Fairness without Demographics.}
\ir{
Several works~\cite{hashimoto2018fairness,ashurst2023fairness,zhao2022towards, yan2020fair,chai2022fairness} have achieved fairness without demographics due to regulatory restrictions.
They generally follow Rawlsian Max-Min fairness~\cite{rawls2017theory} to minimize the risk for the group with the highest risk.
However, these methods depends on unreliable correlations between protected groups and observable features.
Additionally, they mainly focus on general machine learning tasks and may overlook the complexities of user behaviors in RSs.
}

\vspace{-1mm}
\section{Conclusion}
\vspace{-1mm}
In this paper, we improved recommendation fairness without sensitive attributes during training.
We leveraged the capabilities of LLMs to uncover latent sensitive patterns, utilizing them to facilitate fairness training.
Inspired by human reasoning, we employed multi-persona LLMs to mimic diverse human perceptions for sensitive information inference.
We further considered the confusion and consensus among human-like agents to better leverage these inferences.
Comprehensive experiments confirmed the effectiveness of our proposed framework in improving recommendation fairness.

\bibliographystyle{plainnat}
\bibliography{References.bib}

\newpage
\appendix

\section*{Appendix}

\section{Prompts}
Here we provide all the prompts used in the multi-persona sensitive information inference.

\subsection{Persona Editor}
\label{app:persona_editor}
\begin{tcolorbox}[colframe=black, colback=white, boxrule=0.7pt, arc=3pt, width=0.99\columnwidth, boxsep=-0.5mm]
    \small
    \texttt{\small Generate detailed descriptions for {\setlength{\fboxsep}{0.1ex}\colorbox{yellow}{\textit{<number of personas>}}} unique individuals. For each person, include aspects such as cultural background, social norms they adhere to, key personal experiences that shaped their views, and related stereotypes they might encounter, their awareness of current media and trends, psychological traits, age with corresponding generational influences, and their professional field. Provide each characteristic in a concise, clear sentence to depict a well-rounded persona.}
\end{tcolorbox}

\subsection{Annotators with Persona}
\label{app:annotator_with_persona}
\begin{tcolorbox}[colframe=black, colback=white, boxrule=0.7pt, arc=3pt, width=0.99\columnwidth, boxsep=-0.5mm]
    \small
    \texttt{\small \textit{<SYSTEM\_PROMPT>} Imagine you are {\setlength{\fboxsep}{0.1ex}\colorbox{yellow}{\textit{<persona description>}}}. Now you are tasked with analyzing the provided movie viewing history and using your own experience and preference to suggest a likely gender of the user. In your responses, directly specify the inferred gender. Ensure that your responses are concise and avoid redundancy. \textit{</SYSTEM\_PROMPT>}
    \\
    \\
    Here are some movies a user has watched. Infer the user's gender from the movie list.
    \\
    {\setlength{\fboxsep}{0.1ex}\colorbox{yellow}{\textit{<historical behaviors>}}}}
\end{tcolorbox}

\subsection{Meta Summarizer}
\label{app:meta_summarizer}
\begin{tcolorbox}[colframe=black, colback=white, boxrule=0.7pt, arc=3pt, width=0.99\columnwidth, boxsep=-0.5mm]
    \small
    \texttt{\small \textit{<SYSTEM\_PROMPT>} As a meta reviewer, your role is to evaluate the final gender inference based on the user‘s movie viewing history. This involves analyzing other annotators’ inferences and their justifications. Your task is to provide a definitive gender inference along with a comprehensive rationale that synthesizes the insights and reasons provided by the previous annotators. \textit{</SYSTEM\_PROMPT>}
    \\
    \\
    Below are the movie viewing history and the accompanying inferences and justifications from previous annotators.
    \\
    \#\#\# movie viewing history
    \\
    {\setlength{\fboxsep}{0.1ex}\colorbox{yellow}{\textit{<historical behaviors>}}}
    \\
    \#\#\# Annotator 1
    \\
    {\setlength{\fboxsep}{0.1ex}\colorbox{yellow}{\textit{<sensitive attribute inference response>}}}
    \\
    \#\#\# Annotator 2
    \\
    …
    \\
    \\
    Now please deduce the user’s gender. Ensure your inference includes a detailed rationale that integrates and critically assesses the insights and reasoning from the earlier annotations.}
\end{tcolorbox}

\section{Baseline Descriptions}
\label{app:baseline}
we compare \model against several baselines including:
\textit{general fairness methods without demographic information:} 
\textbf{1) ARL}~\cite{lahoti2020fairness} proposes adversarially reweighted learning to improve the utility for worst-off protected groups without access to protected features;
\textbf{2) FKD}~\cite{chai2022fairness} employs knowledge distillation in improving general fairness without observing sensitive attributes;
\textbf{3) EIIL}~\cite{creager2021environment} is a domain-invariant learning approach that incorporates environment inference without using environment labels, where we use FairMI~\cite{zhao2023fair} as the backbone;
\textit{recommendation fairness methods with full sensitive attributes:} 
\textbf{1) ComFair}~\cite{bose2019compositional} uses adversarial filters to remove sensitive information from user and item embeddings;
\textbf{2) FairGo}~\cite{wu2021learning} employs adversarial learning on a user-centric graph to ensure that the sensitive attribute is not exposed by the user's local network;
\textbf{3) FDA}~\cite{chen2023improving} improves fairness from the data augmentation perspective by generating fake interaction to augment the original data;
\textbf{4) FairMI}~\cite{zhao2023fair} defines fairness with a two-fold mutual information objective and utilizes bounds for model optimization;
\textit{recommendation fairness methods with limited or no sensitive attributes:}
\textbf{1) FairGNN}~\cite{dai2021say} introduces a sensitive estimator to predict missing sensitive attributes using graph structures and employs an adversarial approach to enhance fairness;
\textbf{2) FairLISA}~\cite{zhang2023fairlisa} proposes an adversarial framework that utilizes data with both known and unknown sensitive labels;
\textbf{3) AFSR}~\cite{zeng2024fair} identifies stereotypical patterns without sensitive attributes by training gradients, which are then used to smooth item embeddings.

\begin{table}[t]
    \centering
    \renewcommand\arraystretch{1}
     \tabcolsep 0.05in
    \caption{Mathematical Notations.}
    \vspace{-2.5mm}
      \scalebox{0.85}{
      \begin{tabular}{c|l}
      \toprule
      Notation & Description \\
      \midrule
      $\rep{u}, \rep{v}$ & The item and user embeddings \\
      $a_u$ & The actual sensitive attribute of user $u$ \\
      $\tilde{a}_u^i$ &  The $i$-th annotator's inferred sensitive attribute of user $u$ \\
      $\rep{s}, \rep{p}$ & The sensitive-aware and -blind embeddings \\
      $A, S, P, R, E$ & The stochastic variables \\
      $o_i$ & The persona of the $i$-th annotator \\
      $\mathcal{E}(\cdot), \mathcal{S}(\cdot), \mathcal{P}(\cdot)$ & The collaborative, sensitive and preference encoders \\
      $\rep{F}_i$ & The confusion matrix of the $i$-th annotator \\ 
      \bottomrule
      \end{tabular}%
      }
    \label{tab:notation}
  \end{table}%

\section{Experimental Settings}
\subsection{Datasets}
\label{app:dataset}
\begin{wraptable}{r}{0.5\columnwidth}
    \vspace{-5mm}
    \centering
     \tabcolsep 0.06in
    \caption{Data statistics.}
      \scalebox{0.8}{
        \begin{tabular}{c|cccc}
      \toprule
      Dataset & \# Users & \# Items & \# Interactions & Density \\
      \midrule
      \textbf{\ml} & 6,040 & 3,706 & 1,000,209 & 4.47\% \\
      \textbf{\lastfm} & 53,295 & 22,392 & 2,387,940 & 0.20\% \\
      \bottomrule
      \end{tabular}}%
      \vspace{-3mm}
    \label{tab:dataset}
  \end{wraptable}
For the \ml\footnote{\href{https://grouplens.org/datasets/movielens/}{https://grouplens.org/datasets/movielens/}} dataset, we consider ratings greater than 0 as positive feedback.
For the \lastfm\footnote{\href{http://ocelma.net/MusicRecommendationDataset/lastfm-360K.html}{http://ocelma.net/MusicRecommendationDataset/lastfm-360K.html}} dataset, 
we first remove users without profile information and then randomly sample 20\% users of the remaining users to form a subset.
Following the 10-core setting, we also exclude users with fewer than 10 interactions and items with fewer than 10 plays.
The interactions of each user are split into a training set, a validation set, and a test set in an 80\%/10\%/10\% ratio.
The data statistics are shown in \cref{tab:dataset}.

\subsection{Fairness Evaluation Protocol}
\label{app:fairness_evaluation}
For recommendation fairness, following~\cite{zhang2023fairlisa,bose2019compositional,wu2021learning}, we train an attacker to identify sensitive information.
Specifically, users are split into training and test sets in an 80\%/20\% ratio. 
We use the fair embeddings and sensitive attribute labels in the training set to train the attacker, and report the AUC results of sensitive attribute classification on the test set.
If an attacker is able to identify sensitive attributes within fair embeddings, this suggests that such sensitive information is revealed.
Therefore, lower AUC values indicate better fairness with less sensitive information leakage.

\section{Implementation Details}
\label{app:implementation}
In \model, we adopted the open-sourced Llama-3-8B-Instruct\footnote{\href{https://huggingface.co/meta-llama/Meta-Llama-3-8B-Instruct}{https://huggingface.co/meta-llama/Meta-Llama-3-8B-Instruct}} model as the LLM backbone to create the agents. The study on different LLM backbones is also provided in \cref{app:llm_study}.
The text embedding model was chosen as the jina-embeddings-v2-base-en\footnote{\href{https://huggingface.co/jinaai/jina-embeddings-v2-base-en}{https://huggingface.co/jinaai/jina-embeddings-v2-base-en}} model.
As for hyper-parameters, the dimensionality of all latent representations was set to 64.
The batch size for BPR training and sensitive representation learning were set to 2048 and 128, respectively.
The learning rate was set to 0.001.
An additional item-side MI lower bound was optimized due to superior practical performance. 
The hyperparameters $\lambda_{ub}$ and $\lambda_{lb}$ in \cref{eq:bound} were set to 0.01 and 0.1, respectively.
The $\alpha$ in \cref{eq:lb} was set to 0.1.
We also performed a random search for a few hyperparameters on the validation set, including the number of nearest neighbors $K$ (\cref{eq:similarity}) in $\{1,2,3\}$, the number of annotators $N_o$ in $\{4, 6, 8, 10\}$, and the coefficients (\cref{eq:sens_overall}) $\lambda_{sim}$ and $\lambda_{fine}$ in $\{10^{-2}, 10^{-3}, 10^{-4}\}$.
The sensitive information inference module was conducted on NVIDIA A40-48GB GPUs.
The sensitive representation learning module was conducted on NVIDIA RTX 3090 GPUs.


\begin{figure}[t]
    \centering
    \begin{minipage}{0.55\columnwidth}
        \centering
        \includegraphics[width=0.99\columnwidth]{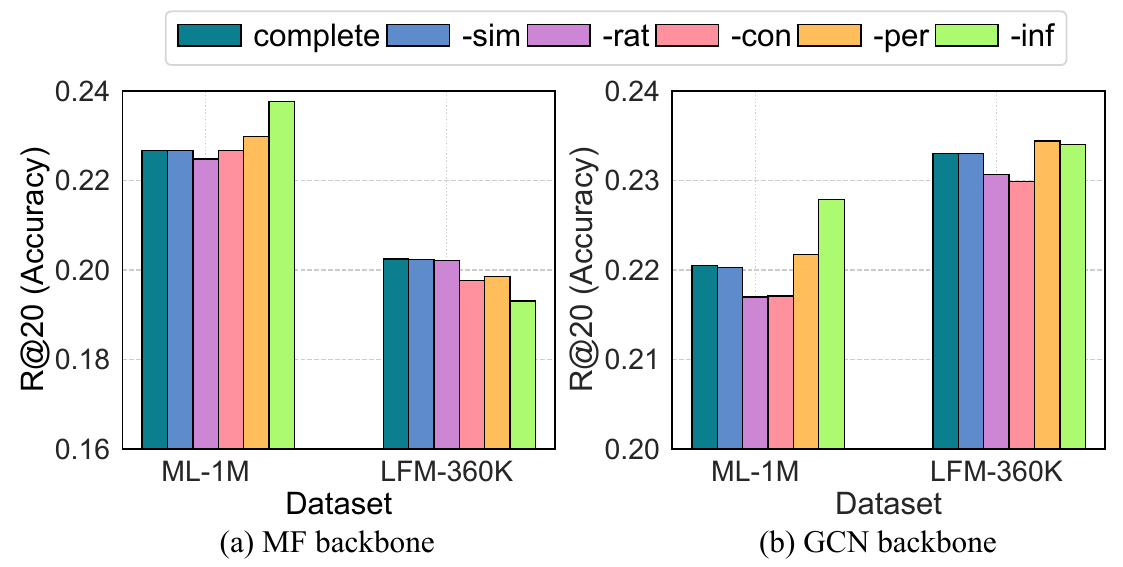}
        \vspace{-1mm}
        \caption{
            Contributions of different components in \model to the recommendation accuracy.
        }
        \vspace{-2.5mm}
        \label{fig:app_ablation}
    \end{minipage}
    \hspace{1ex}
    \begin{minipage}{0.42\columnwidth}
        \centering
      \subfigure[\ml, $N_o$]{
          \includegraphics[width=0.45\columnwidth]{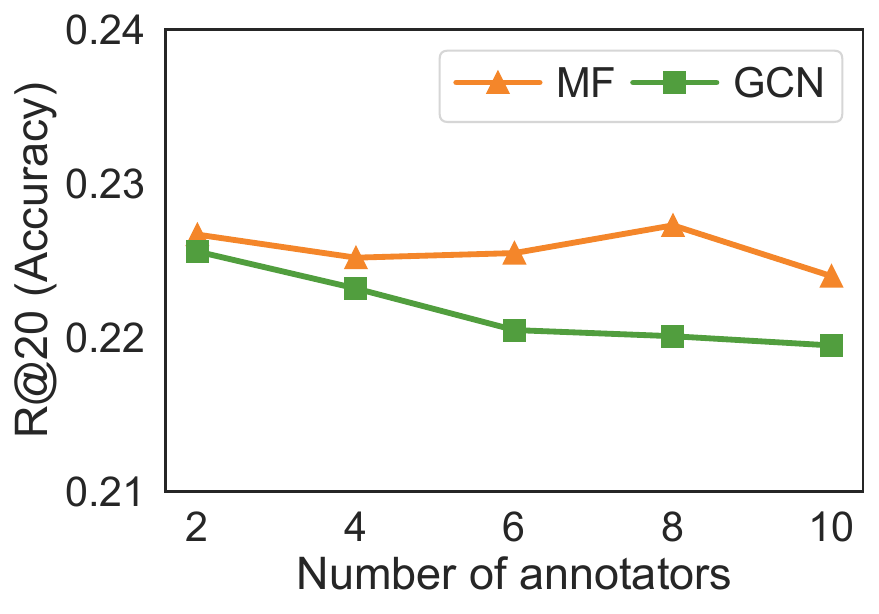}
              \label{fig:sens_num_ml_rec}
      }\hspace{-1mm}
      \subfigure[\lastfm, $N_o$]{
          \includegraphics[width=0.45\columnwidth]{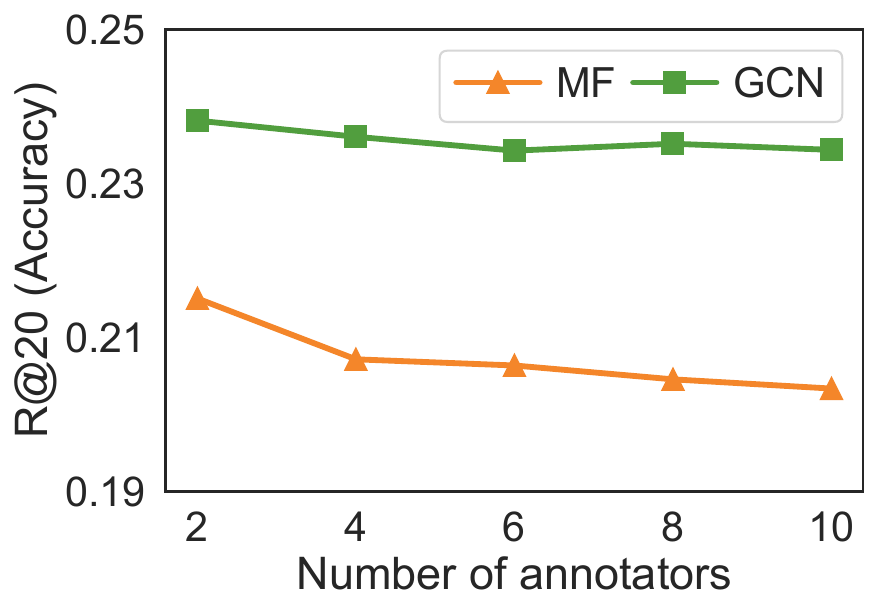}
              \label{fig:sens_num_lastfm_rec}
      }\vspace{-3mm}
      \subfigure[\ml, $\lambda_{sim}$]{
          \includegraphics[width=0.45\columnwidth]{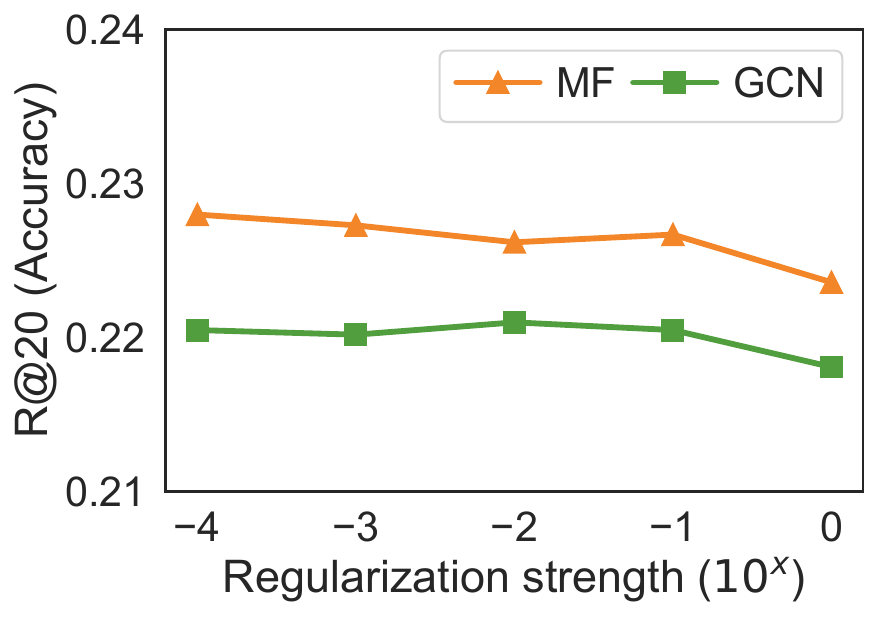}
              \label{fig:sens_strength_ml_rec}
      }\hspace{-1mm}
      \subfigure[\lastfm, $\lambda_{sim}$]{
          \includegraphics[width=0.45\columnwidth]{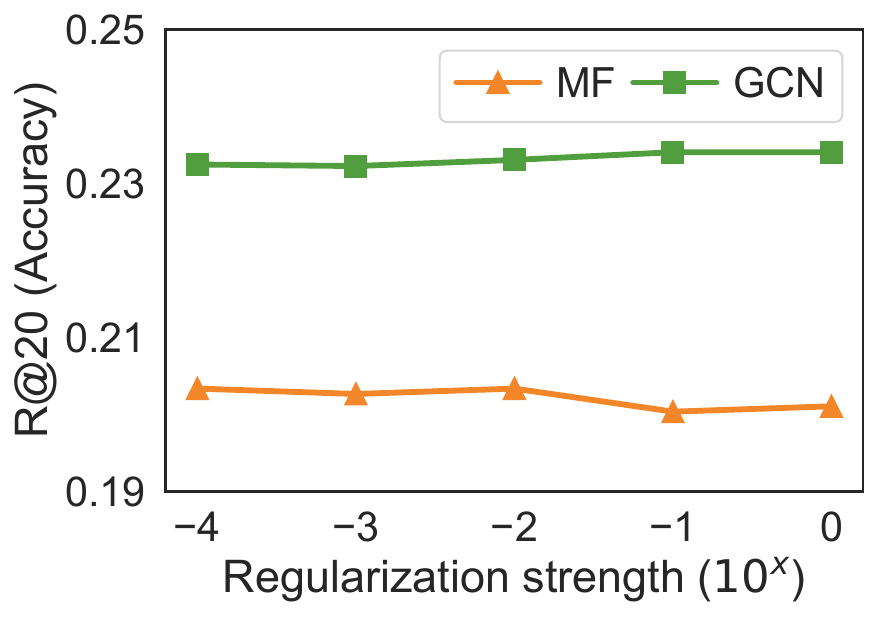}
              \label{fig:sens_strength_lastfm_rec}
      }
        \vspace{-4mm}
        \caption{Parameter sensitivity of recommendation accuracy on hyperparameters.}
        \label{fig:app_sens}
    \end{minipage}
\end{figure}

\section{Additional Experimental Results}
\label{app:exp}
\subsection{Ablation and Sensitivity Analysis on Recommendation Accuracy}
\label{app:ablation_sensitivity}
We present the recommendation accuracy (R@20) results for the analysis of ablation study and parameter sensitivity in \cref{fig:app_ablation,fig:app_sens}, respectively.
Overall, we observe an inverse relationship between recommendation accuracy and fairness. This is expected, as these two considerations often conflict by nature.
Importantly, the results indicate that variations in recommendation accuracy remain within an acceptable range.

\subsection{LLM Backbone Study}
\label{app:llm_study}
\begin{wrapfigure}{r}{0.5\columnwidth}
    \vspace{-6mm}
    \centering
    \subfigure[MF backbone]{
        \includegraphics[width=0.23\columnwidth]{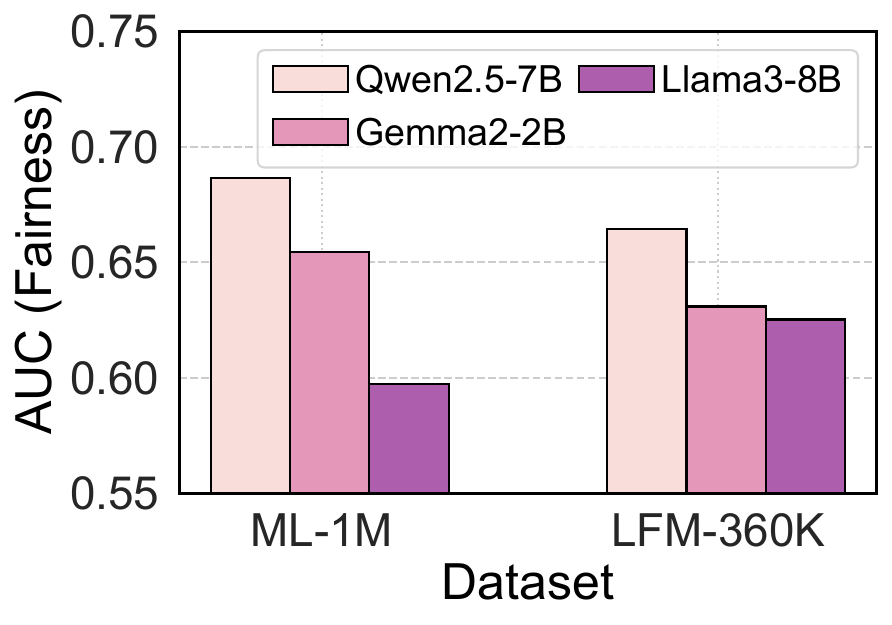}
    }\hspace{-1mm}
    \subfigure[GCN backbone]{
        \includegraphics[width=0.23\columnwidth]{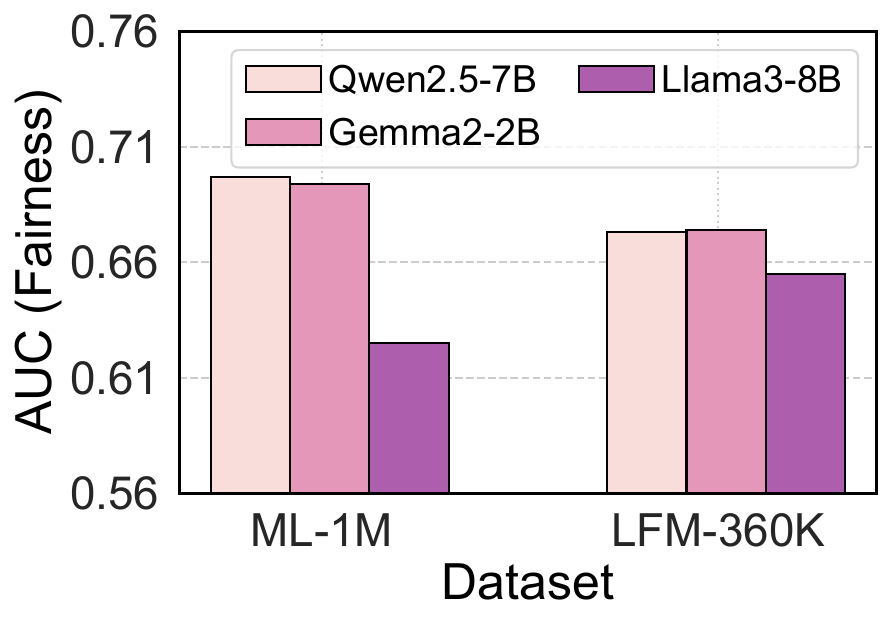}
    }
    \vspace{-3mm}
    \caption{The fairness of \model when equipped with different LLM backbones.}
    \vspace{-5mm}
    \label{fig:llm}
  \end{wrapfigure}
  We also evaluate the performance of LLMFOSA using different LLM backbones. Specifically, we experiment with two widely used open-source LLMs, \ie Qwen2.5-7B\footnote{\href{https://huggingface.co/Qwen/Qwen2.5-7B-Instruct}{https://huggingface.co/Qwen/Qwen2.5-7B-Instruct}} and Gemma2-2B\footnote{\href{https://huggingface.co/google/gemma-2-2b-it}{https://huggingface.co/google/gemma-2-2b-it}}, alongside Llama3-8B, which is used as the primary backbone in this study.
  As shown in \cref{fig:llm}, \model with Llama3-8B achieves the best fairness performance, likely due to its larger parameter size and well-documented reasoning capabilities. This justifies our choice of Llama3-8B as the backbone for this work. 
  Furthermore, Gemma2-2B outperforms Qwen2.5-7B in most cases despite having fewer parameters, likely because its training on a large English corpus provides an advantage for English-based inference tasks.

\subsection{Performance on Occupation Attribute}
\label{app:occupation}

\begin{table}[t]
    \centering
    \renewcommand\arraystretch{1.1}
   \tabcolsep 0.07in
  \caption{The overall fairness (AUC) and recommendation accuracy (N@20 and R@20) performance on Occupation attribute. "-" indicates that the method is not applicable to the mutiple attribute setting.}
    \scalebox{0.75}{
      \begin{tabular}{c|>{\columncolor[rgb]{0.898,0.898,0.898}}c|cc|>{\columncolor[rgb]{0.898,0.898,0.898}}c|cc}
      \toprule
      \textit{Backbone} & \multicolumn{3}{c|}{MF} & \multicolumn{3}{c}{GCN} \\
      \midrule
      \textit{Metrics} & AUC($\downarrow$)  & R($\uparrow$) & N($\uparrow$) & AUC($\downarrow$)  & R($\uparrow$) & N($\uparrow$) \\
      \midrule
      NoFair & 0.6652 & 0.2288 & 0.2313 & 0.6617 & 0.2363 & 0.2376 \\
      \midrule
      ARL  & 0.6750 & 0.2284 & 0.2305 & 0.6699 & 0.2438 & 0.2436 \\
      FKD  & 0.6842 & 0.2243 & 0.2256 & 0.6566 & 0.2402 & 0.2451 \\
      EIIL & 0.6691 & 0.2212 & 0.2297 & 0.6781 & 0.2344 & 0.2342 \\
      \midrule
      ComFair & 0.6717 & 0.2117 & 0.2190 & 0.6583 & 0.2085 & 0.2152 \\
      FairGo & 0.6600 & 0.2011 & 0.2084 & 0.6563 & 0.2012 & 0.2094 \\
      FDA  & - & - & - & - & - & - \\
      FairMI & 0.6609 & 0.2285 & 0.2283 & 0.6648 & 0.2228 & 0.2283 \\
      \midrule
      FairGNN & 0.6632 & 0.2141 & 0.2142 & 0.6597 & 0.2179 & 0.2189 \\
      FairLISA & 0.6585 & 0.1952 & 0.1965 & 0.6534 & 0.2046 & 0.2094 \\
      AFSR & 0.6528 & 0.1438 & 0.1572 & 0.6519 & 0.2295 &0.2351 \\
      \midrule
      \model & \textbf{0.6385} & 0.2278 & 0.2326 & \textbf{0.6513} & 0.2380 & 0.2412 \\
      \bottomrule
      \end{tabular}}%
    \label{tab:occupation}%
  \end{table}%

To further assess the generalizability of \model beyond the gender attribute, we extend our evaluation to the Occupation attribute in the \ml dataset, considering 21 distinct occupation categories.
As shown in \cref{tab:occupation}, \model demonstrates competitive fairness performance relative to existing methods, supporting the broader applicability of our approach.
Notably, although several baselines achieve fairness levels comparable to \model when using the GCN backbone (\eg AFSR, FairLISA, FairGo), they suffer from a substantial decline in accuracy, which limits their practicality for real-world deployment.

\section{Bound Derivations}
\subsection{Maximizing $I(S;A)$}
\label{app:a_bound}
Recall that we use the annotators' inferred sensitive attributes to introduce $\widetilde{A}$, an approximation of $A$, which takes values in the inferred sensitive attribute labels. Then, based on the definition of MI, we have:
\begin{equation}
    \label{eq:mi_coarse_approx}
        I(S;A) \doteq \mathbb{E}_{(\rep{s}_u, \tilde{a}_u^i) \sim p(S,\widetilde{A})}\left[\log{\frac{p(\tilde{a}_u^i|\rep{s}_u)}{p(\tilde{a}_u^i)}}\right],
    \end{equation}
    where $\tilde{a}^i_u$ is the inferred sensitive attribute of user $u$ by annotator agent $i$.
    As $p(\tilde{a}_u^i|\rep{s}_u)$ is intractable, we introduce the variational distribution $q(\tilde{a}_u^i|\rep{s}_u)$:
    \begin{equation}
    \label{eq:variational}
        {I}(S;A)
        \doteq \mathbb{E}_{(\rep{s}_u, \tilde{a}_u^i) \sim p(S,\widetilde{A})}
        \left[
        \log \frac{q_(\tilde{a}_u^i|\rep{s}_u)}{p(\tilde{a}_u^i)}
        \right] + \mathbb{E}_{\tilde{a}_u^i \sim p(\widetilde{A})} 
        \left[
        \mathrm{KL}\left( p(\tilde{a}_u^i|\rep{s}_u) \| q(\tilde{a}_u^i | \rep{s}_u) \right)
        \right].
    \end{equation}
Based on \cref{eq:variational}, we maximize $I(S;A)$ by maximizing the following lower bound:
\begin{equation}
	\label{eq:app:sens_bound}
		\max_{\mathcal{S}, \mathcal{C}} \mathbb{E}_{(\rep{s}_u, \tilde{a}_u^i) \sim p(S,\widetilde{A})}
		\left[
		 \log q(\tilde{a}_u^i|\rep{s}_u; \mathcal{C})
		\right],
	\end{equation}
    where $\mathcal{C}(\cdot)$ is the classifier used to classify the sensitive label based on the sensitive-aware embedding, also implemented as an MLP.

\subsection{Minimizing $I(S;P)$}
\label{app:mi_ub}
We derive an upper bound of $I(S;P)$ so as to effectively minimize the MI. We start by the proposition as follows:
\begin{prop}
    \textit{Given $\rep{s} \sim p(\rep{s}_u)$}, if the conditional distribution $p(\rep{s}_u | \rep{p}_u)$ is known, then the MI $I(S;P)$ can be upper bounded by:
    \begin{equation}
        I(S;P) \le \mathbb{E}\left[ \log p(\rep{s}_u | \rep{p}_u) - \frac{1}{M} \sum_{u^\prime \in \mathcal{U}} \log p(\rep{s}_{u^\prime} | \rep{p}_u) \right],
    \end{equation}
\end{prop}
The proof can be found in \cite{cheng2020club}.
Afterwards, we utilize a variational distribution $q(\rep{s}_u | \rep{p}_u)$ to approximate the conditional distribution $p(\rep{s}_u | \rep{p}_u)$ by minimizing the KL divergence:
\begin{equation}
    \min_{q(\cdot)} \text{KL}\left(p(\rep{s}_u | \rep{p}_u) \Vert q(\rep{s}_u | \rep{p}_u)\right).
\end{equation}
To simplify the optimization, we assume that the variational distribution $q(\rep{s}_u | \rep{p}_u)$ is a Gaussian distribution and the loss function (\cref{eq:loss_ub}) for the MI minimization can be derived:
\begin{equation}
    \mathcal{L}_{ub} = \sum_{u \in \mathcal{U}} \left[
    \log \mathcal{B}(\rep{p}_u) - \frac{1}{M} \sum_{u^\prime \in \mathcal{U}} \log \mathcal{B}(\rep{p}_{u^\prime})
    \right],
\end{equation}
where $\mathcal{B}(\cdot)$ is a neural network to parameterize the distribution $q(\rep{s}_u | \rep{p}_u)$.

\subsection{Maximizing $I(P;R|S)$}
\label{app:mi_lb}
We derive another lower bound of $I(P;R|S)$ to maximize the MI. We first introduce the following proposition:
\begin{prop}
    \textit{Given $\rep{r}_u, \rep{p}_u, \rep{s}_u \sim p(\rep{r}_u, \rep{p}_u, \rep{s}_u), \rep{r}^\prime \sim p(\rep{r}_u|\rep{s}_u)$ and a score function $f(\cdot)$, the MI $I(P;R|S)$ can be lower bounded by:}
    \begin{equation}
        I(P;R|S) \ge \mathbb{E}\left[\log \frac{\exp f(\rep{r}_u, \rep{p}_u, \rep{s}_u)}{\frac{1}{M} \sum_{u^\prime \in \mathcal{U}} \exp f(\rep{r}_{u^\prime}, \rep{p}_u, \rep{s}_u)}\right],
    \end{equation}
\end{prop}
The proof can be found in \cite{gupta2021controllable}. Then, the loss function for the MI maximization can be derived:
\begin{equation}
\mathcal{L}_{lb} = 
- \sum_{u \in \mathcal{U}} \left[
    \log \frac{\exp f(\rep{r}_u, \rep{p}_u, \rep{s}_u)}{\frac{1}{M} \sum_{u^\prime \in \mathcal{U}} \exp f(\rep{r}_{u^\prime}, \rep{p}_u, \rep{s}_u)}
\right].
\end{equation}
For the implementation, we take the score function $f(\cdot)$ as the weighted cosine similarity: $f(\rep{r}_u, \rep{p}_u, \rep{s}_u) = \langle \rep{r}_u, \rep{p}_u + \alpha \rep{s}_u \rangle$, where $\alpha$ is a hyperparameter.

\section{Dicsussion on Applicability}
\label{app:applicability}
In our proposed \model framework, the multi-persona inference phase involves only lightweight inference operations, without the need for heavy fine-tuning, and is executed prior to the fairness training stage. Since this phase can be performed offline, it does not interfere with the primary fairness training process or the real-time recommendation generation, which rely solely on a lightweight recommendation model. This design ensures that LLMFOSA remains a practical and feasible solution for real-world applications, even with its initial computational requirements.

\section{Illustrative Responses from LLM Agents}
\label{app:response}
To better demonstrate the inference capabilities of LLM agents in the proposed Multi-Persona Sensitive Information Inference module, we randomly select one response each from an annotator and the meta summarizer. These responses are shown in \cref{fig:response_annotator,fig:response_meta}, respectively.
As evident from these examples, LLM-powered agents can infer users' sensitive attributes based on their historical behaviors and summarize the rationales for these inferences.
These abilities ensure the reliable discovery of sensitive patterns without relying on explicitly provided sensitive attributes.

\begin{figure}[t]
    \centering
\begin{tikzpicture}
    \node[anchor=text,text width=0.9\columnwidth, draw, sharp corners, line width=0.8pt, fill=white, inner sep=2mm, align=justify] (big) {
        \begin{spacing}{0.85}
        \vspace{-1ex}
     { \small Based on the movie viewing history and the accompanying inferences and justifications from the previous annotators, I infer that the user is likely a \textbf{female}. Here's my \textbf{comprehensive rationale}:
\\
\\
1. The presence of classic Disney movies like Snow White, Cinderella, and Beauty and the Beast is a common thread throughout the annotations. These films are \textbf{often associated with a female audience} and suggest a fondness for traditional fairy tales and romance.
\\
2. The inclusion of romantic comedies and dramas, such as My Fair Lady, Meet Joe Black, and Titanic, is another consistent theme. \textbf{These genres often appeal to women} and suggest an interest in relationships and emotional storytelling.
\\
3. The presence of family-friendly movies like Toy Story, James and the Giant Peach, and Mulan, as well as animated films like Aladdin, Hercules, and Bambi, \textbf{suggests a love for lighthearted, feel-good stories and a willingness to engage with popular culture.}
\\
4. \textbf{The absence of action-oriented or sci-fi movies, which are often popular among male audiences}, is a notable pattern. This suggests that the user may be less interested in these genres and more inclined towards character-driven, emotionally resonant storytelling.
\\
5. \textbf{The presence of strong female protagonists in films} like Erin Brockovich, Mulan, and Run Lola Run, as well as the user's interest in romantic musicals like Gigi, My Fair Lady, and The Sound of Music, \textbf{suggests an appreciation for female empowerment and strong female characters.}
\\
6. The overall tone of the movie list is characterized by a focus on romance, relationships, and family-friendly entertainment, which is \textbf{often associated with a female audience.}
\\
\\
While it's possible that a male user could have similar tastes, the consistency of these themes and patterns across the annotations suggests that the user is likely a female. The user's movie list is characterized by a strong interest in traditional fairy tales, romance, and family-friendly entertainment, \textbf{which are all common interests among women.}
\\
\\
In conclusion, based on the movie viewing history and the accompanying inferences and justifications from the previous annotators, \textbf{I infer that the user is likely a female}. The user's movie list is characterized by a focus on romance, relationships, and family-friendly entertainment, and the presence of strong female protagonists and classic Disney movies suggests an appreciation for female empowerment and traditional fairy tales.}
        \vspace{-2.5ex}
        \end{spacing}
    };
\end{tikzpicture}
\vspace{-2.5mm}
\caption{An example of the meta summarizer's response.}
\label{fig:response_meta}
\end{figure}

\begin{figure}[t]
        \centering
    \begin{tikzpicture}
        \node[anchor=text,text width=0.9\columnwidth, draw, sharp corners, line width=0.8pt, fill=white, inner sep=2mm, align=justify] (big) {
        	\begin{spacing}{0.85}
        	\vspace{-1ex}
         { \small Based on the movie list, I infer that the user is likely \textbf{male}.
\\
\\
The list includes a mix of classic and popular films, but with a noticeable bias towards action, adventure, and comedy movies that are often appealing to a male audience. The presence of Indiana Jones, Crocodile Dundee, Blade, and Men in Black suggests a fondness for action-packed films. The inclusion of Animal House, Young Guns II, and Dances with Wolves also \textbf{suggests a preference for masculine-themed movies.}
\\
\\
While there are some films that might appeal to a female audience, such as The Princess Bride, Little Mermaid, and The Full Monty, the overall tone and genre suggests a \textbf{male} user.}
         	\vspace{-2.5ex}
        	\end{spacing}
        };
    \end{tikzpicture}
    \vspace{-2.5mm}
    \caption{An example of the annotator's response.}
    \label{fig:response_annotator}
\end{figure}

\section{Limitations and Broader Impact}
\label{app:limitations}

Although the proposed LLMFOSA framework achieves promising fairness improvements, it has some limitations.
First, the sensitive inference depends on the capabilities of the selected LLMs, and choosing an appropriate model may require some effort.
Second, the confusion modeling is based on the common assumption that the confusion is independent of the user's behaviors given the ground-truth sensitive attribute. In the future, we plan to explore the instance-dependent transition matrix to enhance the confusion modeling.

Recommender systems significantly influence users' daily decisions.
In this paper, we discuss the social responsibilities of recommender systems and consider a situation where no sensitive attributes are available.
We contribute to the development of recommender systems that are both ethically sound and practically deployable in privacy-conscious environments.

\end{document}